\documentclass{article}
\usepackage[preprint]{spconf}
\usepackage{spconf,amsmath,amssymb,graphicx}
\usepackage[numbers,sort&compress]{natbib}

\usepackage{booktabs}
\usepackage[safe]{tipa}
\usepackage{microtype}
\usepackage[hidelinks]{hyperref}
\usepackage{eso-pic}

\begin{document}
\ninept
\title{Does a Prosody-Trained Representation Help Beyond Trainable Fusion? A Parameter-Matched Study with Frozen HuBERT}
%
\name{Ki Woong Moon$^{1\star}$ \qquad Daniel Brenner$^{2}$ 
\thanks{\hspace{-1.8em}\normalsize $^{\star} $ Corresponding author.}}

\address{$^{1}$Department of Linguistics, University of Arizona, Tucson, AZ 85721 \\
$^{2}$Speak, 360 Spear St., 4F, San Francisco, CA 94105}

%
\maketitle

\copyrightnotice{\parbox{\textwidth}{This work has been submitted to the IEEE for possible publication.\\ Copyright may be transferred without notice, after which this version may no longer be accessible.}}

\begin{abstract}
Explicit prosodic cues may help automatic speech recognition (ASR) of spontaneous speech, but auxiliary representations typically require additional trainable components, making it unclear whether gains come from the auxiliary information or the fusion mechanism. We address this using a frozen HuBERT backbone and a 64-dimensional representation trained to predict $\log F_0$, voicing, $\Delta\log F_0$, log energy, and spectral tilt. We compare a frozen-backbone recognizer (Baseline), trainable fusion with zero auxiliary input (Null), and the same fusion supplied with the learned representation (Learned). Across Buckeye, Switchboard, and AMI IHM, Null reduces WER by 0.71--1.45 points over Baseline, whereas Learned differs from Null by $+0.07$, $-0.09$, and $+0.00$ points, with no significant differences. However, removing or mismatching the representation at inference increases Learned WER. Thus, Learned depends on the representation yet shows no measurable incremental WER benefit over the parameter-matched control.
\end{abstract}


\section{Introduction}
\label{sec:intro}

Spontaneous speech is acoustically variable and ambiguous, as segments may be shortened, weakened, deleted, or heavily coarticulated. For example,  ``I don't know'' can be realized as \textipa{[\~{a} \~{o}n\~{o}]}, with substantial deletion and coarticulation. Johnson \citep{johnson2004massive} reported that such reduction in spontaneous speech is systematic, with roughly a quarter of words in conversational American English containing at least one segment deletion. The degree of reduction also varies with linguistic predictability and prosodic prominence, with more predictable materials tending to be shorter and less phonetically prominent \citep{aylett2006language, turnbull2017role}. Acoustic measures associated with prominence include F0, loudness, duration, and spectral balance \citep{turnbull2017role, sluijter1996spectral, kochanski2005loudness}. Such cues may provide useful information for recognizing acoustically ambiguous spontaneous speech, raising the question of how effectively modern speech representations capture and exploit them.

Self-supervised learning (SSL) models such as HuBERT \citep{hsu2021hubert} learn useful representations directly from the audio signal. SSL representations have also been shown to encode prosodically relevant information \citep{lin2023, de2023prosaudit}. However, recoverability does not establish that a downstream ASR system effectively exploits this information during recognition.

Yet evaluating the contribution of an auxiliary representation introduces an attribution problem. Incorporating auxiliary information typically requires additional trainable components, such as projections, adapters, or fusion modules. A comparison between a frozen SSL baseline and an auxiliary-conditioned model therefore changes two factors simultaneously: the information supplied to the recognizer and the trainable mechanism used to incorporate it. Any recognition improvement in such a comparison cannot be attributed uniquely to the auxiliary information. This confound is particularly relevant when the backbone is frozen, because the added trainable mechanism provides an additional pathway for transforming otherwise fixed SSL representations.

To separate the effects of auxiliary information from those of the trainable fusion mechanism used to incorporate it, we use a two-phase system. We first train a compact prosody encoder to predict $\log F_0$, voicing, $\Delta\log F_0$, $\log$ energy, and spectral tilt, yielding a 64-dimensional prosody-trained representation. In the ASR system, we keep this representation frozen and use it to condition trainable fusion modules applied to the saved transformer-layer states of a frozen HuBERT. Importantly, we compare the \textit{Learned} condition that receives the learned auxiliary representation with the \textit{Null} condition that has an identical fusion architecture but receives a zero-valued auxiliary input. Because the two conditions have the same architecture and number of trainable parameters, their comparison tests the incremental contribution of the auxiliary representation while holding the fusion architecture and trainable parameter count fixed. The effect of adding the trainable fusion pathway is evaluated by comparing the \textit{Null} with \textit{Baseline}, which omits the fusion modules.

With this design, we ask whether adding the trainable fusion pathway improves recognition, whether the learned representation provides additional benefit beyond a parameter-matched zero-input pathway, and whether the trained \textit{Learned} model uses the representation at inference. Across three corpora, the trainable pathway consistently improves recognition, whereas the learned representation provides little additional WER improvement. We use inference-time interventions to test the last question directly.\footnote{\normalsize
Code and reproducibility materials are available \href{https://github.com/Ki-Woong95/prosody-or-adaptation}{here}.}

\begin{figure*}[!t]
    \centering
    \includegraphics[width=\textwidth,
    ]{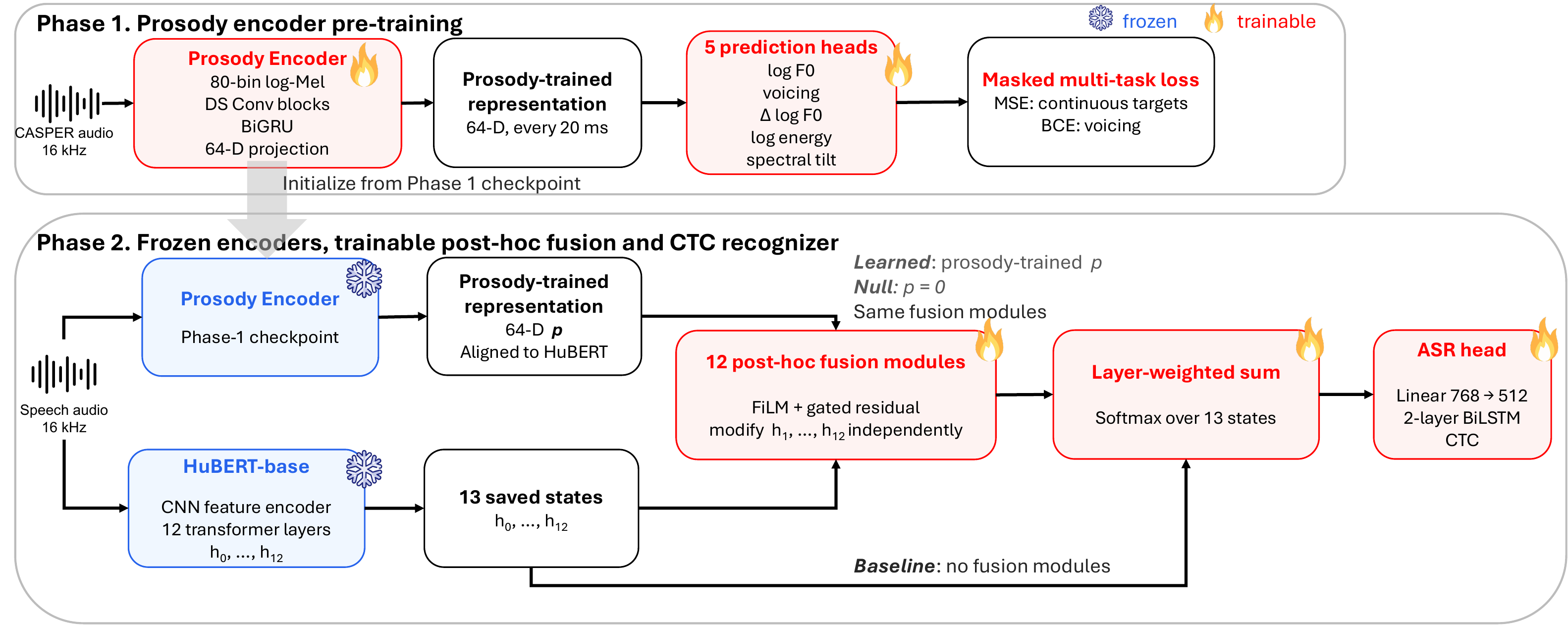}
    \caption{Two-phase system. Phase~1 trains a 64-D representation using five acoustic-prosodic targets. In Phase~2, frozen HuBERT and the frozen Phase~1 encoder produce representations independently. Twelve post-hoc fusion modules modify $h_1,\ldots,h_{12}$; $h_0$ bypasses fusion, and modified states are not fed back into HuBERT. Baseline omits fusion, Null uses the same modules with $p=0$, and Learned uses the prosody-trained $p$.}
    \vspace{-0.40cm}
    \label{fig:model_architecture}
\end{figure*}
\section{Related Work}
\label{sec:backgrounds}

Phonetic realization in spontaneous speech varies systematically with lexical predictability and prosodic prominence \citep{jurafsky2001effect, bell2009predictability, aylett2006language, turnbull2017role}. This motivates interest in whether pretrained speech representations encode information relevant to such variation. Layerwise analyses show that acoustic and linguistic information is distributed non-uniformly across SSL encoder layers \citep{pasad2023comparative}. Prosodically relevant information is also recoverable from these representations: SSL models support prosody reconstruction and future-prosody prediction \citep{lin2023} and distinguish strong from weak prosodic boundaries \citep{de2023prosaudit}. However, recoverability does not imply that a downstream ASR system effectively exploits the same information during recognition. 

Recent work has explored explicit prosodic supervision in pretrained speech recognizers. Sasu and Schluter \citep{sasu2025pitch}, for example, jointly train pitch-accent detection and ASR with wav2vec~2.0 \citep{baevski2020wav2vec} and report improved recognition performance. Their comparison contrasts ASR-only training with a joint auxiliary objective, leaving open whether the improvement is specific to the prosodic supervision or reflects the broader effect of introducing the auxiliary training objective. Our study addresses a related but distinct attribution question by holding the trainable fusion mechanism fixed while varying whether it receives an informative auxiliary representation.

The difficulty of attributing downstream performance to a frozen representation rather than to its trainable head is well established in probing. Hewitt and Liang \citep{hewitt2019control} introduced control tasks to contextualize probe accuracy, and Zaiem et al.\ \citep{zaiem2025probing} showed that changing the probing head can alter speech SSL model rankings. Related ASR systems also inject auxiliary cues through trainable pathways, including FiLM conditioning on enhanced speech \citep{yang2022film}. Such systems couple the auxiliary information with a trainable mechanism for incorporating it. Our \textit{Learned}--\textit{Null} comparison instead holds the fusion pathway fixed while varying its auxiliary input; the interventions separately test whether the trained system uses the representation.

\section{Methods}
\label{sec:methods}

\subsection{Overview}
The system tests the contributions of the prosody-trained representation and the trainable fusion mechanism in two phases. 
Phase~1 trains a compact encoder to produce a frame-level representation supervised by five acoustic-prosodic targets. Phase~2 uses this representation to condition trainable fusion modules applied independently to the hidden-state output of each HuBERT transformer layer. Because each modified state is used only for downstream layer aggregation and is not passed to the next HuBERT layer, we call this operation post-hoc layerwise fusion. Figure~\ref{fig:model_architecture} summarizes the system.

\subsection{Data}
Phase~1 uses CASPER \citep{xiao2025casper}. We resample recordings to 16~kHz and divide them into non-overlapping segments of at most 15~s. We use a fixed 80/20 segment-level training/validation split. CASPER is used only to train and select the Phase~1 encoder and is disjoint from the Phase~2 ASR corpora.

Phase~2 uses three English corpora spanning conversational and meeting settings with different recording conditions. Buckeye \citep{pitt2005buckeye} contains close-microphone spontaneous speech, Switchboard \citep{godfrey1992sbd} contains two-party telephone conversations, and AMI~IHM \citep{kraaij2005ami} contains multi-party meeting speech recorded with individual headset microphones. Buckeye uses 30/5/5 speaker-disjoint train/validation/test speakers, corresponding to 4,915/797/925 segments and 26.86/4.35/5.06~h.  For Switchboard, we follow the preprocessing and train/validation/test splits of Ho et al. \citep{ho2025swb} (185,402/20,601/51,501 utterances) and remove annotations enclosed in angle or square brackets. For AMI~IHM, the train/validation/test partitions contain 75,174/9,428/8,514 usable utterances.

\subsection{Phase 1: Prosody-trained encoder}

The Phase~1 encoder maps 16~kHz waveforms to a 64-dimensional representation at a 20~ms frame interval. Its frontend computes an 80-bin log-mel spectrogram using a 50~ms window, 20~ms hop, and non-centered framing. The log-mel features are projected to 128 dimensions, then passed through four depthwise-separable convolutional blocks \citep{howard2017mobilenets}, a bidirectional GRU \citep{cho2014gru}, and finally projected to 64 dimensions.

Five prediction heads predict $\log F_0$, voicing, $\Delta\log F_0$, log energy, and spectral tilt. $F_0$ and periodicity are estimated using CREPE-tiny \citep{kim2018crepe} and resampled by timestamp to the encoder frame grid, with $F_0$ restricted to 50--500~Hz.  A frame is treated as voiced when its energy exceeds the utterance-specific 20th percentile, periodicity is at least .01, and $F_0$ lies within the valid range. In the Phase~1 training set, the periodicity criterion excluded 22.1\% of frames that passed the energy criterion. $\Delta\log F_0$ is defined only across consecutive voiced frames, and continuous targets are standardized using statistics estimated from eligible frames in the Phase~1 training partition. Targets undefined at a given frame are masked from the corresponding loss rather than imputed.

Mean-squared error is used for the continuous targets and binary cross-entropy for voicing. The training objective is
\begin{equation}
L =
L_{\log F_0}
+ L_{\mathrm{voi}}
+ 0.5L_{\Delta \log F_0}
+ 0.5L_{\mathrm{eng}}
+ 0.5L_{\mathrm{tilt}},
\end{equation}

where each term is averaged over eligible, non-padded frames. Phase~1 uses a batch size of 32, learning rate of $10^{-3}$, and early stopping (patience 10) for up to 50 epochs, selecting the checkpoint with the lowest masked validation loss. Only the resulting 64-D hidden representation is retained for Phase~2.

\subsection{Phase~2: Post-hoc HuBERT fusion}

We use HuBERT-base \citep{hsu2021hubert} with the SSL backbone frozen throughout Phase~2. HuBERT produces a pre-transformer hidden state $h_0$ followed by twelve transformer-layer outputs $h_1, \ldots, h_{12}$, each with 768 dimensions. For the fusion conditions, the frozen Phase~1 encoder is evaluated independently, and the non-padded representation for each utterance is linearly resampled to that utterance's HuBERT frame length before fusion.

For each transformer layer $\ell\geq1$, let $h_\ell$ denote the frozen HuBERT state and $p\in\mathbb{R}^{B\times T\times64}$ the aligned auxiliary representation. The fusion module applies FiLM-style conditioning \citep{perez2018film} followed by a gated residual:
\begin{align}
\bar p &= \mathrm{LayerNorm}(p), \\
(\gamma,\beta) &= \mathrm{Linear}(\bar p),\\
c &= \mathrm{LayerNorm}(h_\ell+\gamma\odot h_\ell+\beta),\\
g &= \sigma\!\left(\mathrm{Linear}([h_\ell;p])\right),\\
h_{\mathrm{out},\ell}
&= h_\ell+\tanh(s_\ell)\,g\odot(c-h_\ell),
\end{align}

where $\gamma,\beta\in\mathbb{R}^{B\times T\times768}$, $g\in\mathbb{R}^{B\times T\times1}$ is broadcast along the feature dimension, $\odot$ denotes element-wise multiplication, and $[h_\ell;p]$ denotes concatenation along the feature dimension. Each fusion module's learnable residual scale $s_\ell$ is initialized to zero, making the module an exact identity mapping at initialization. State $h_0$ bypasses the fusion modules. A learned softmax-weighted sum over all thirteen states is then projected from 768 to 512 dimensions, passed through a two-layer bidirectional LSTM \citep{graves2005bidirectional}, and fed into a CTC output layer \citep{graves2006CTC}.

\subsection{Experimental conditions and training}

We compare three conditions. \textit{Baseline} uses frozen HuBERT without the post-hoc fusion modules. \textit{Null} includes all twelve fusion modules, with the auxiliary input set to zero throughout training and inference. \textit{Learned} uses the identical fusion architecture with the learned auxiliary representation as input. \textit{Null} and \textit{Learned} therefore have identical architectures and trainable parameter counts (12.16M), whereas \textit{Baseline} has 10.93M trainable parameters. Although matched in trainable parameter count, \textit{Null} cannot exploit utterance-varying auxiliary conditioning because its auxiliary input is fixed to zero.

Only the layer-aggregation weights, post-hoc fusion modules when present, 768-to-512 projection, BiLSTM, and CTC head are optimized. HuBERT remains frozen and in evaluation mode throughout training; the Phase~1 encoder is also frozen and kept in evaluation mode when present. Training uses AdamW \citep{loshchilov2019decoupled} with a learning rate of \(10^{-4}\), weight decay of 0.01, and \((\beta_1,\beta_2)=(0.9,0.98)\). We use a batch size of 8 with gradient accumulation over 4 steps (effective batch size 32), mixed-precision training, and gradient clipping at 1.0. Early stopping is based on validation WER with a patience of 10 epochs. We use greedy CTC decoding for evaluation. We use three seeds per corpus and condition.

\subsection{Statistical analysis}

Our primary comparison is \textit{Learned}--\textit{Null}, which tests whether supplying the learned auxiliary representation improves recognition relative to the same trainable fusion mechanism with its auxiliary input fixed to zero. \textit{Null}--\textit{Baseline} is a secondary comparison assessing the effect of adding the trainable fusion pathway.

Statistical inference is based on paired test-set predictions and a hierarchical Poisson bootstrap with 100,000 draws. In each bootstrap draw, training seeds are resampled while preserving the pairing between conditions. For Buckeye and AMI~IHM, speakers are also resampled before applying Poisson weights to utterances within each speaker. Switchboard lacks usable speaker labels in the stored predictions and therefore uses seed-utterance resampling. We report absolute WER percentage-point differences, 95\% confidence intervals, and two-sided $p$-values. Holm correction is applied separately across the three corpus-level \textit{Learned}--\textit{Null} tests and the three \textit{Null}--\textit{Baseline} tests.
\begin{table*}[t]
\centering
\caption{Test WER (\%, mean $\pm$ SD over three training seeds) with 95\% hierarchical bootstrap CIs for absolute WER differences. Negative values favor the first condition; bold indicates the lowest unrounded mean WER.}
\label{tab:main_results}

\begin{tabular}{lccccc}
\toprule
Corpus &
Baseline &
Null &
Learned &
Learned$-$Null &
Null$-$Baseline \\
\midrule
Buckeye &
37.27 $\pm$ 0.10 &
\textbf{35.82} $\pm$ 0.13 &
35.89 $\pm$ 0.11 &
+0.07 [$-$0.22, +0.34] &
$-$1.45 [$-$1.93, $-$0.94] \\

Switchboard &
31.95 $\pm$ 0.09 &
31.24 $\pm$ 0.04 &
\textbf{31.15} $\pm$ 0.07 &
$-$0.09 [$-$0.21, +0.03] &
$-$0.71 [$-$0.85, $-$0.58] \\

AMI IHM &
39.11 $\pm$ 0.22 &
\textbf{38.23} $\pm$ 0.22 &
38.23 $\pm$ 0.20 &
+0.00 [$-$0.27, +0.28] &
$-$0.89 [$-$1.24, $-$0.54] \\
\bottomrule
\end{tabular}
\vspace{-0.15cm}
\end{table*}

\subsection{Inference-time diagnostics}

To determine whether the trained \textit{Learned} model makes functional use of its auxiliary representation, each seed-specific \textit{Learned} checkpoint is evaluated without retraining under four input conditions: the original representation (\textsc{True}), an all-zero representation (\textsc{Zero}), a within-utterance time-shuffled representation (\textsc{Time}), and a representation obtained from a different utterance (\textsc{Utt}). For \textsc{Utt}, donor utterances are assigned deterministically by a circular shift of half the test set, without constraining speaker identity, and the donor representation is resampled to the recipient utterance's HuBERT frame length.

These interventions test the dependence of an already-trained \textit{Learned} model on its auxiliary representation and must be distinguished from the \textit{Null} training condition. \textsc{Utt} substitutes a representation from another utterance, \textsc{Time} disrupts its temporal alignment, and \textsc{Zero} removes the learned input entirely. Because \textit{Learned} is never trained with an all-zero auxiliary sequence, \textsc{Zero} is an out-of-distribution sanity check rather than a matched control. \textit{Null}, by contrast, is optimized from initialization with zero auxiliary input. Exploratory intervention contrasts use the same paired hierarchical Poisson bootstrap with 100,000 draws, with Holm correction applied jointly across the nine corpus-by-intervention contrasts.

Finally, we quantify each fusion module's residual magnitude
$r_{\ell,t} =
\tanh(s_\ell)g_{\ell,t}\odot(c_{\ell,t}-h_{\ell,t})$
relative to the corresponding frozen HuBERT state:
\begin{equation}
\rho_\ell =
\mathbb{E}_{t\in\mathrm{valid}}
\left[
\frac{\|r_{\ell,t}\|_2}
{\|h_{\ell,t}\|_2}
\right].
\end{equation}
We report these layerwise quantities descriptively rather than performing post-hoc significance tests across individual layers.

\section{Results}
\label{sec:results}
\subsection{Recognition performance}

To assess cross-corpus agreement of the frozen Phase~1 encoder, we evaluated its target predictions on the CASPER validation set and the three downstream validation sets. On CASPER, correlations for $\log F_0$, $\Delta\log F_0$, log energy, and spectral tilt were .889, .703, .998, and .998, respectively, and voicing F1 was .890. Across Buckeye, Switchboard, and AMI~IHM, the corresponding correlations were .804/.793/.944 for $\log F_0$, .694/.786/.830 for $\Delta\log F_0$, .996/.993/.988 for log energy, and .997/.995/.996 for spectral tilt; voicing F1 was .882/.864/.880. These diagnostics measure agreement with automatically extracted supervision targets rather than independent ground-truth prosody.

Table~\ref{tab:main_results} reports test WER averaged over three training seeds. \textit{Learned} differs from \textit{Null} by $+0.069$ WER percentage points on Buckeye, $-0.090$ on Switchboard, and $+0.004$ on AMI~IHM. All three confidence intervals include zero, and none is significant after Holm correction (all $p_{\mathrm{Holm}}\geq .400$). Moreover, the 95\% confidence intervals are compatible with at most a few tenths of a percentage point of improvement from the learned representation under the present architecture.

In contrast, \textit{Null} consistently outperforms \textit{Baseline} by 0.71--1.45 WER points. All confidence intervals exclude zero ($p_{\mathrm{Holm}}<.001$), with the same direction for every seed. The trainable fusion pathway therefore yields a robust improvement, whereas the learned representation adds little further benefit.

\subsection{Functional use of the auxiliary representation}

The near-zero \textit{Learned}--\textit{Null} difference does not imply that \textit{Learned} ignores its auxiliary input. We therefore intervene on the input at inference without retraining; Table~\ref{tab:interventions} reports the WER increase relative to the unmodified model.

\begin{table}[t]
\centering
\caption{WER increase (percentage points) after intervening on the auxiliary representation of the trained Learned model. Positive values indicate degradation relative to the unmodified representation.}
\label{tab:interventions}

\begin{tabular}{lrrr}
\toprule
Corpus & Time shuffle & Utt.\ subst. & Zero \\
\midrule
Buckeye     & +0.17 & +0.42 & +4.85 \\
Switchboard & +0.30 & +1.06 & +16.93 \\
AMI IHM     & +0.35 & +1.01 & +12.77 \\
\bottomrule
\end{tabular}
\vspace{-0.15cm}
\end{table}

Utterance substitution increased WER by 0.42--1.06 points and was significant on all three corpora (all $p_{\mathrm{Holm}} \leq .0061$), indicating use of utterance-specific auxiliary information. Temporal shuffling produced smaller increases of 0.17--0.35 points, significant for Switchboard and AMI~IHM but not Buckeye, suggesting weaker sensitivity to temporal ordering. Replacing the representation with zeros produced larger degradations of 4.85--16.93 points; because this input is out of distribution for \textit{Learned}, we treat it as a sanity check rather than direct evidence of incremental utility.

\subsection{Fusion behavior}
\begin{figure}[t]
    \centering
    \includegraphics[width=\columnwidth]{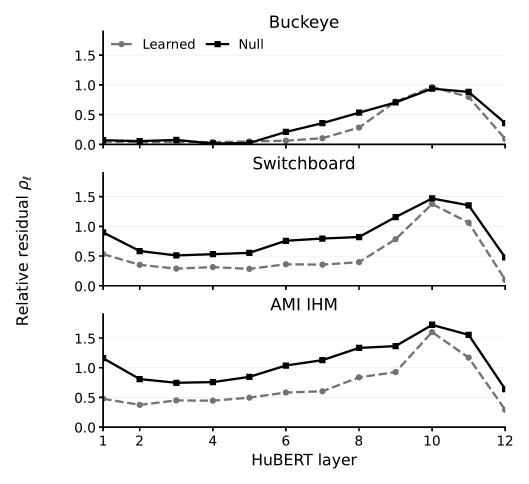}
    \caption{Layerwise relative residual magnitude \(\rho_\ell\), averaged over three seeds. Larger values indicate greater modification of frozen HuBERT states.}
    \label{fig:layer_residuals}
\vspace{-0.20cm}
\end{figure}

Both \textit{Null} and \textit{Learned} modify frozen HuBERT states, especially in upper layers (Figure~\ref{fig:layer_residuals}). Mean relative residual magnitude is larger for \textit{Null} than \textit{Learned} on every corpus: .351 versus .249 on Buckeye, .825 versus .526 on Switchboard, and 1.091 versus .672 on AMI~IHM. The nonzero residuals show that \textit{Null} is an active control, while \textit{Learned} shows its strongest modifications around layers 9--11.

\section{Discussion}
\label{sec:discussion}

The trainable fusion mechanism improves the frozen HuBERT recognizer, whereas the prosody-trained representation yields little additional WER reduction. Under this architecture, most of the observed gain is associated with the trainable fusion pathway rather than the auxiliary representation.

\textit{Learned} nevertheless depends on its auxiliary input: utterance substitution degrades recognition even though \textit{Null} achieves comparable WER without an utterance-varying signal. One possible explanation is redundancy: frozen HuBERT states may already encode information correlated with the supervised acoustic-prosodic cues, leaving little incremental information for the auxiliary representation. 

These conclusions are limited to one SSL backbone, one supervision set that omits duration, a CTC recognizer, and a frozen post-hoc design. Full fine-tuning, propagating fused states through HuBERT, or localized error analysis may yield different effects. Parameter-matched controls remain important for separating gains due to auxiliary information from those due to the trainable mechanism used to incorporate it.

\clearpage

\section{Compliance with Ethical Standards}
This research study was conducted retrospectively using previously collected human speech data from the CASPER \citep{xiao2025casper}, Buckeye \citep{pitt2005buckeye}, Switchboard \citep{godfrey1992sbd}, and AMI \citep{kraaij2005ami} corpora. No funding was received for conducting this study. The authors have no relevant financial or nonfinancial interests to disclose.

\bibliographystyle{IEEEbib}
\bibliography{refs}

@inproceedings{hewitt2019control,
  author    = {Hewitt, John and Liang, Percy},
  title     = {Designing and Interpreting Probes with Control Tasks},
  booktitle = {EMNLP-IJCNLP},
  pages     = {2733--2743},
  year      = {2019}
}

@article{zaiem2025probing,
  author  = {Zaiem, Salah and Kemiche, Youcef and Parcollet, Titouan and Essid, Slim and Ravanelli, Mirco},
  title   = {Speech Self-Supervised Representations Benchmarking: A Case for Larger Probing Heads},
  journal = {Comput. Speech Lang.},
  volume  = {89},
  pages   = {101695},
  year    = {2025}
}

@inproceedings{yang2022film,
  author    = {Yang, Da-Hee and Chang, Joon-Hyuk},
  title     = {{FiLM} Conditioning with Enhanced Feature to the Transformer-Based End-to-End Noisy Speech Recognition},
  booktitle = {Interspeech},
  pages     = {4098--4102},
  year      = {2022}
}

@article{hsu2021hubert,
  author  = {Hsu, Wei-Ning and Bolte, Benjamin and Tsai, Yao-Hung Hubert and Lakhotia, Kushal and Salakhutdinov, Ruslan and Mohamed, Abdelrahman},
  title   = {{HuBERT}: Self-Supervised Speech Representation Learning by Masked Prediction of Hidden Units},
  journal = {IEEE/ACM Trans. Audio, Speech, Lang. Process.},
  volume  = {29},
  pages   = {3451--3460},
  year    = {2021}
}

@inproceedings{baevski2020wav2vec,
  author    = {Baevski, Alexei and Zhou, Yuhao and Mohamed, Abdelrahman and Auli, Michael},
  title     = {wav2vec 2.0: A Framework for Self-Supervised Learning of Speech Representations},
  booktitle = {NeurIPS},
  year      = {2020}
}

@inproceedings{lin2023,
  author    = {Lin, Guan-Ting and Feng, Chi-Luen and Huang, Wei-Ping and others},
  title     = {On the Utility of Self-Supervised Models for Prosody-Related Tasks},
  booktitle = {SLT},
  year      = {2022}
}

@inproceedings{de2023prosaudit,
  author    = {de Seyssel, Maureen and Lavechin, Marvin
               and Titeux, Hadrien and others},
  title     = {{ProsAudit}: A Prosodic Benchmark for
               Self-Supervised Speech Models},
  booktitle = {Interspeech},
  pages     = {2963--2967},
  year      = {2023},
}

@inproceedings{sasu2025pitch,
  author    = {Sasu, David and Schluter, Natalie},
  title     = {Pitch Accent Detection Improves Pretrained Automatic Speech Recognition},
  booktitle = {Interspeech},
  year      = {2025}
}

@inproceedings{pasad2023comparative,
  author    = {Pasad, Ankita and Shi, Bowen and Livescu, Karen},
  title     = {Comparative Layer-wise Analysis of Self-Supervised Speech Models},
  booktitle = {ICASSP},
  year      = {2023}
}

@inproceedings{kim2018crepe,
  author    = {Kim, Jong Wook and Salamon, Justin and Li, Peter and Bello, Juan Pablo},
  title     = {{CREPE}: A Convolutional Representation for Pitch Estimation},
  booktitle = {ICASSP},
  year      = {2018}
}

@inproceedings{cho2014gru,
  author    = {Cho, Kyunghyun and van Merri{\"e}nboer, Bart and Gulcehre, Caglar and others},
  title     = {Learning Phrase Representations Using {RNN} Encoder--Decoder for Statistical Machine Translation},
  booktitle = {EMNLP},
  year      = {2014}
}

@inproceedings{graves2005bidirectional,
  author    = {Graves, Alex and Fern{\'a}ndez, Santiago and Schmidhuber, J{\"u}rgen},
  title     = {Bidirectional {LSTM} Networks for Improved Phoneme Classification and Recognition},
  booktitle = {ICANN},
  year      = {2005}
}

@article{howard2017mobilenets,
  author  = {Howard, Andrew G. and Zhu, Menglong and Chen, Bo and others},
  title   = {{MobileNets}: Efficient Convolutional Neural Networks for Mobile Vision Applications},
  journal = {arXiv preprint arXiv:1704.04861},
  year    = {2017}
}

@inproceedings{perez2018film,
  author    = {Perez, Ethan and Strub, Florian and de Vries, Harm and Dumoulin, Vincent and Courville, Aaron},
  title     = {{FiLM}: Visual Reasoning with a General Conditioning Layer},
  booktitle = {AAAI},
  year      = {2018}
}

@inproceedings{loshchilov2019decoupled,
  author    = {Loshchilov, Ilya and Hutter, Frank},
  title     = {Decoupled Weight Decay Regularization},
  booktitle = {ICLR},
  year      = {2019}
}

@inproceedings{graves2006CTC,
  author    = {Graves, Alex and Fern{\'a}ndez, Santiago and Gomez, Faustino and Schmidhuber, J{\"u}rgen},
  title     = {Connectionist Temporal Classification: Labelling Unsegmented Sequence Data with Recurrent Neural Networks},
  booktitle = {ICML},
  year      = {2006}
}

@inproceedings{xiao2025casper,
  title={{CASPER}: A large scale spontaneous speech dataset},
  author={Xiao, Cihan and Liang, Ruixing and Zhang, Xiangyu and Tiryaki, Mehmet Emre and Bae, Veronica and Shankar, Lavanya and Yang, Rong and Poon, Ethan and Dupoux, Emmanuel and Khudanpur, Sanjeev and others},
  booktitle={2025 IEEE Automatic Speech Recognition and Understanding Workshop (ASRU)},
  pages={1--7},
  year={2025},
  organization={IEEE}
}

@article{pitt2005buckeye,
  title={The {Buckeye} corpus of conversational speech: labeling conventions and a test of transcriber reliability},
  author={Mark A. Pitt and Keith Johnson and Elizabeth Hume and Scott F. Kiesling and William D. Raymond},
  journal={Speech Commun.},
  year={2005},
  volume={45},
  pages={89-95},
}

@inproceedings{godfrey1992sbd,
  author    = {Godfrey, John J. and Holliman, Edward C. and McDaniel, Jane},
  title     = {{SWITCHBOARD}: Telephone Speech Corpus for Research and Development},
  booktitle = {ICASSP},
  year      = {1992}
}

@inproceedings{kraaij2005ami,
  author    = {McCowan, I. and Carletta, J. and Kraaij, W. and Ashby, S. and
               Bourban, S. and Flynn, M. and Guillemot, M. and Hain, T. and
               Kadlec, J. and Karaiskos, V. and Kronenthal, M. and Lathoud, G. and
               Lincoln, M. and Lisowska, A. and Post, W. and Reidsma, Dennis and
               Wellner, P.},
  title     = {The {AMI} Meeting Corpus},
  booktitle = {Measuring Behavior},
  year      = {2005}
}

@inproceedings{ho2025swb,
  title     = {{Enhancing Transcripts of Open-Source Automatic Speech Recognition Models Through Fine-Tuning with Laughter and Speech-Laugh}},
  author    = {Phuoc Hoang Ho and Dragoș Alexandru Bălan and Dirk K. J. Heylen and Khiet P. Truong},
  year      = {2025},
  booktitle = {{Interspeech}},
  pages     = {4513--4517},
  doi       = {10.21437/Interspeech.2025-2193},
  issn      = {2958-1796},
}

@incollection{johnson2004massive,
  author    = {Johnson, Keith},
  title     = {Massive Reduction in Conversational {American} {English}},
  booktitle = {Spontaneous Speech: Data and Analysis},
  editor    = {Yoneyama, Kiyoko and Maekawa, Kikuo},
  pages     = {29--54},
  publisher = {National Institute for Japanese Language},
  address   = {Tokyo, Japan},
  year      = {2004}
}

@article{aylett2006language,
  author  = {Aylett, Matthew and Turk, Alice},
  title   = {Language Redundancy Predicts Syllabic Duration and the Spectral Characteristics of Vocalic Syllable Nuclei},
  journal = {JASA},
  volume  = {119},
  number  = {5},
  pages   = {3048--3058},
  year    = {2006}
}

@article{bell2009predictability,
  author  = {Bell, Alan and Brenier, Jason M. and Gregory, Michelle and Girand, Cynthia and Jurafsky, Dan},
  title   = {Predictability Effects on Durations of Content and Function Words in Conversational {English}},
  journal = {J. Mem. Lang.},
  volume  = {60},
  number  = {1},
  pages   = {92--111},
  year    = {2009}
}

@inproceedings{jurafsky2001effect,
  author    = {Jurafsky, Daniel and Bell, Alan and Gregory, Michelle and Raymond, William D.},
  title     = {The Effect of Language Model Probability on Pronunciation Reduction},
  booktitle = {ICASSP},
  year      = {2001}
}

@article{sluijter1996spectral,
  author  = {Sluijter, Agaath M. C. and van Heuven, Vincent J.},
  title   = {Spectral Balance as an Acoustic Correlate of Linguistic Stress},
  journal = {JASA},
  volume  = {100},
  number  = {4},
  pages   = {2471--2485},
  year    = {1996}
}

@article{turnbull2017role,
  author  = {Turnbull, Rory},
  title   = {The Role of Predictability in Intonational Variability},
  journal = {Lang. Speech},
  volume  = {60},
  number  = {1},
  pages   = {123--153},
  year    = {2017}
}

@article{kochanski2005loudness,
  author  = {Kochanski, Greg and Grabe, Esther and Coleman, John and Rosner, Burton},
  title   = {Loudness Predicts Prominence: Fundamental Frequency Lends Little},
  journal = {JASA},
  volume  = {118},
  number  = {2},
  pages   = {1038--1054},
  year    = {2005}
}

\end{document}